\documentclass[proceedings]{JHEP}

\conference{Quantum aspects of gauge theories, supersymmetry and unification
 ,Paris France,1-7 September 1999.
}

\title{
Some remarks on the free fields realization of the bosonic string on 
 \( AdS_{3} \).
}

\author{
Igor Pesando \thanks { Preprint DFTT 09/00;
    Work supported partially by the European Commission TMR programme
    ERBFMRX-CT96-0045.} \\
Dipartimento di Fisica Teorica , Universit\'a di Torino\\
Istituto Nazionale di Fisica Nucleare (INFN) - sezione di Torino, Italy \\
via P. Giuria 1, I-10125 Torino\\
 \email{ipesando@to.infn.it} 
}

\abstract{
We discuss the classical solutions of the equations of motion and the possible 
boundary condition for a bosonic string with Kalb-Ramond background 
in \( AdS_{3} \). 
It turns out that there exists three different physical sectors and that it is 
also possible to describe the motion on an extremal black hole background. 
The existence of three sectors clearly shows how one of the spectra proposed is 
incomplete. We consider also the classical 'canonical' transformation
which maps 
the string fields to the classical Wakimoto ones. 
It turns however out that the Wakimoto fields are not free because of the 
boundary conditions and  in order to have the usual mode expansion with 
reasonable behaviour under complex conjugation it is necessary to consider 
the complexification of \( AdS_{3} \) and then add some constraints. 
Furthermore they cover only half \( AdS_{3} \) and we need different patches 
to cover the whole space and to make the above mentioned transformation 
really canonical.
}

\csname @addtoreset\endcsname{equation}{section}
\makeatother

\begin{document}


\section{Introduction.}

The last year has seen an increasing and renewed interest in string theory
propagating on \( AdS \) backgrounds. A particular attention has been dedicated
to the \( AdS_{3} \) background with NSNS flux, this because it is the only
non trivial model which has been possible to treat exactly at the quantum level.
Nevertheless and despite its apparent simplicity our understanding is far from
being complete. In particular the question of which is its spectrum is not satisfactorily
settled. It was already noted in the old days that with a naive quantization
the Virasoro constraints were not sufficient to eliminate the ghost from the
spectrum (\cite{balog}). In order to remedy to this unfortunate circumstance
two different proposals have been put forward\footnote{
Some recent papers have done some steps forwards in increasing our understanding
of the quantum theory (\cite{MaldaOo},\cite{LaSa},\cite{giapu}). Especially
(\cite{MaldaOo}) has given a nice interpretation of the old results by (\cite{Hwang})
on the construction of a modular invariant partition function.
}

\begin{enumerate}
\item To truncate the spectrum to the unitary part: this was the approach first advocated
by Petropolous (\cite{Petro}) and independently by Mohammedi (\cite{Moha}),
developed by Hwang (\cite{Hwang}) and collaborators. Last year Evans, Gaberdiel
and Perry (\cite{Gabe}) showed that the free spectrum is actually ghost free;
\item To introduce some new hidden dof in form of zero momenta of the fields used
to bosonize the KM currents. This way was first undertaken by Bars (\cite{Bars})
and developed by Satoh (\cite{Satoh}) along slightly different lines.
\end{enumerate}
We think that both proposals have some weak sides but the main criticism is
due to their philosophical attitude to the problem (see also (\cite{Marios})
for a recent review of the open problems): string theory is so consistent that
there should not be necessary to introduce new elements in the game since the
string itsself should give the answer.

Our approach is different from both the previous ones even if it is closer to
the second one: we find the classical canonical transformation from 
the string
fields to the Wakimoto fields and then we discuss how and in what extent the
Wakimoto fields are free.

\section{The Classical Bosonic String Theory on \protect\( AdS_{3}\protect \).}

We first exam the equations of motion and boundary conditions for a classical
bosonic string with a Kalb-Ramond background thought of as a WZW theory plus
Virasoro constraints not relying on the KM symmetry structure present in the
theory and we derive the general solution.
We set up the canonical formalism which we use to discuss the classical
'canonical' transformation to  the classical Wakimoto fields. 
finally we
discuss the boundary conditions and their consequences.
It turns out that
the Wakimoto fields while satisfying free fields equations of motion are generically
not free because of the boundary conditions; we discuss how to circumvent the
problem.

\subsection{The action and the constraints of the bosonic string on \protect\( AdS_{3}\protect \).}

The string action is given by\footnote{
\begin{itemize}
\item Spacetime coordinates: \( x^{\pm }=\frac{x^{1}\pm x^{0}}{\sqrt{2}} \)
\item WS metric \( -\eta _{00}=\eta _{11}=1 \); \( \eta _{+-}=\frac{1}{2} \) \( \eta ^{+-}=2 \) 
\item WS coordinates: \( \xi ^{\pm }=\tau \pm \sigma  \); \( \sigma \in [0,2\pi ] \);
\( d^{2}\xi =2d\tau d\sigma  \); \( \partial _{\pm }=\frac{1}{2}\left( \partial _{\tau }\pm \partial _{\sigma }\right)  \) 
\item Sigma matrices: \( \sigma _{3}=\left( \begin{array}{cc}
1 & \\
 & -1
\end{array}\right)  \)\( \sigma _{+}=\left( \begin{array}{cc}
 & 1\\
0 & 
\end{array}\right)  \)\( \sigma _{-}=\left( \begin{array}{cc}
 & 0\\
1 & 
\end{array}\right)  \)
\item \( sl(2,R) \) algebra: \( \sigma _{a}\sigma _{b}=\epsilon _{abc}\sigma ^{c}+\eta _{ab}1 \)
with \( \epsilon _{+-3}=\frac{1}{2} \) and \( \eta _{+-}=\frac{1}{2} \) \( \eta _{33}=1 \)\end{itemize}
}

\begin{eqnarray}
S_{wzw}&=&\frac{|k|}{4\pi }\int d^{2}\xi \: tr(\omega _{+}\omega _{-})
\nonumber \\
&&+\frac{k}{12\pi }\int \: tr(\omega ^{3})  \label{Swzw} 
\end{eqnarray}
where \( \omega =g^{-1}dg=\omega _{+}d\xi ^{+}+\omega _{-}d\xi ^{-} \) is the
pullback on the string worldsheet of the left invariant one form on the group
\( SL(2,R) \). When we use the explicit expression for \( g \) in the Gauss
form\footnote{
This expression only covers half \( AdS_{3} \) , even if we perform an analytic
continuation letting \( e^{\rho }\in R \) they do not cover the whole manifold
since points like \( \left( \begin{array}{cc}
\alpha  & \beta \\
-\frac{1}{\beta } & 0
\end{array}\right)  \) are left out.
} 
\begin{equation}
\label{g-gauss}
g=\left( \begin{array}{cc}
1 & \\
x^{-} & 1
\end{array}\right) \left( \begin{array}{cc}
e^{\rho } & \\
 & e^{-\rho }
\end{array}\right) \left( \begin{array}{cc}
1 & x^{+}\\
 & 1
\end{array}\right) =\left( \begin{array}{cc}
e^{\rho } & e^{\rho }x^{+}\\
e^{\rho }x^{-} & e^{-\rho }+e^{\rho }x^{+}x^{-}
\end{array}\right) 
\end{equation}
and take \( k=-|k| \) the previous expression for the action becomes
\begin{equation}
\label{S$}
S_{wzw}=\frac{|k|}{2\pi }\int d^{2}\xi \: \partial _{+}\rho \partial _{-}\rho +e^{2\rho }\partial _{-}x^{+}\partial _{+}x^{-}
\end{equation}
We can interpret this as a string action in the conformal gauge if we add the
Virasoro constraints
\begin{eqnarray*}
T_{++} & = & |k|\left[ \left( \partial _{+}\rho \right) ^{2}+e^{2\rho }\partial _{+}x^{+}\partial _{+}x^{-}\right] =0\\
T_{--} & = & |k|\left[ \left( \partial _{-}\rho \right) ^{2}+e^{2\rho }\partial _{-}x^{+}\partial _{-}x^{-}\right] =0
\end{eqnarray*}
 We want now to derive the the equations of motion and the allowed boundary
conditions associated with the action (\ref{S$}) in the same way of we proceed
with the usual string action in Minkowski space. The equations of motion read
\begin{eqnarray}
\partial _{-}\left( e^{2\rho }\partial _{+}x^{-}\right)  
=  \partial _{+}\left( e^{2\rho }\partial _{-}x^{+}\right) &=&0
\label{eq-mot-x-} \\
\partial _{-}\partial _{+}\rho +e^{2\rho }\partial_{-}x^{+}\partial_{+}x^{-}  
& = & 0\label{eq-mot-rho} 
\end{eqnarray}
while from the request of the cancellation of the surface terms obtained from
the fields variation we get the boundary conditions
\begin{eqnarray}
\delta \rho |_{\sigma =0}=\delta \rho |_{\sigma =2\pi } & \Rightarrow&
\nonumber\\
\rho '|_{\sigma =0}&=&\rho '|_{\sigma =2\pi }
\label{bc-rho} 
\nonumber\\
\delta x^{-}|_{\sigma =0}=\delta x^{-}|_{\sigma =2\pi } & \Rightarrow  & 
\nonumber\\
  e^{2\rho }\partial _{-}x^{+}|_{\sigma =0}&=&e^{2\rho }\partial _{-}x^{+}|_{\sigma =2\pi }\label{bc-x-} \\
\delta x^{+}|_{\sigma =0}=\delta x^{+}|_{\sigma =2\pi } & \Rightarrow& 
\nonumber\\
e^{2\rho }\partial _{+}x^{-}|_{\sigma =0}&=&e^{2\rho }\partial _{+}x^{-}|_{\sigma =2\pi }\label{bc-x+} 
\end{eqnarray}

Anticipating the expressions for the KM currents (\ref{J-},\ref{Jbar3}) we
can rewrite the last two conditions (\ref{bc-x-},\ref{bc-x+}) as 

\begin{eqnarray*}
J^{-}|_{\sigma =0}=J^{-}|_{\sigma =2\pi }\: \: \:  & \overline{J}^{+}|_{\sigma =0}=\overline{J}^{+}|_{\sigma =2\pi } & 
\end{eqnarray*}

\subsection{The general solution of the equations of motions.}

We are now ready to discuss the general solution of eq.s (\ref{eq-mot-x-}-\ref{eq-mot-rho})
. We can write the general solution as
\begin{eqnarray}
x^{+} & = & a(\xi ^{+})+e^{-2c(\xi ^{+})}\frac{\overline{b}(\xi ^{-})}{1+\overline{b}(\xi ^{-})b(\xi ^{+})}\label{sol_x+} \\
x^{-} & = & \overline{a}(\xi ^{-})+e^{-2\overline{c}(\xi ^{-})}\frac{b(\xi ^{+})}{1+\overline{b}(\xi ^{-})b(\xi ^{+})}\label{sol_x-} \\
\rho  & = & \lg \left( 1+\overline{b}(\xi ^{-})b(\xi ^{+})\right) 
\nonumber\\
&&+c(\xi ^{+})+\overline{c}(\xi ^{-}).\label{sol_rho} 
\end{eqnarray}
 from the knowledge of the general solution of the equations of motion associated
with a WZW action, i.e\footnote{
The reason why we choose such a parametrization is because we want the canonical
momenta associated with \( x^{\pm } \) be symmetric in the exchange of the
barred and unbarred quantities.
}.
\begin{eqnarray}
g(\xi ^{+},\xi ^{-}) & = & g^{T}_{R}(\xi ^{-})g_{L}(\xi ^{+})\label{g_grgl} \\
g^{T}_{R}(\xi ^{-}) & = & \left( \begin{array}{cc}
e^{\overline{c}} & e^{\overline{c}}\overline{b}\\
e^{\overline{c}}\overline{a} & e^{\overline{c}}\overline{a}\overline{b}+e^{-\overline{c}}
\end{array}\right) 
\nonumber\\
g_{L}(\xi ^{+}) &=& \left( \begin{array}{cc}
e^{c} & e^{c}a\\
e^{c}b & e^{c}ab+e^{-c}
\end{array}\right) \label{gr} 
\end{eqnarray}

As it is well known the solution (\ref{g_grgl}) does not fix completely \( g_{R},g_{L} \)
(\ref{gr}) which are determined up to a redefinition 
\begin{equation}
\label{gauge-inv}
g_{L}\rightarrow g_{0}g_{L}\: \: \: g_{R}\rightarrow g_{0}^{-T}g_{R}
\end{equation}
 we can (partially) fix this invariance by choosing a canonical form for the
monodromies. This invariance is also connected to the possibility of using different
charts: our parametrization is not global and therefore the group has to be
covered with charts where one patch is parametrized as in (\ref{gr}) and the
others can be chosen to be 
\begin{eqnarray*}
g_{(1)L}(\xi ^{+})&=&\left( \begin{array}{cc}
e^{c_{(1)}}b_{(1)} & e^{c_{(1)}}a_{(1)}b_{(1)}+e^{-c_{(1)}}\\
-e^{c_{(1)}} & -e^{c_{(1)}}a_{(1)}
\end{array}\right) 
\\
g_{(3)L}(\xi ^{+})&=&-\left( \begin{array}{cc}
e^{c_{(3)}}b_{(3)} & e^{c_{(3)}}a_{(3)}b_{(3)}+e^{-c_{(3)}}\\
-e^{c_{(3)}} & -e^{c_{(3)}}a_{(3)}
\end{array}\right)
\\ 
g_{(2)L}(\xi ^{+})&=&-\left( \begin{array}{cc}
e^{c_{(2)}} & e^{c_{(2)}}a_{(2)}\\
e^{c_{(2)}}b_{(2)} & e^{c_{(2)}}a_{(2)}b_{(2)}+e^{-c_{(2)}}
\end{array}\right)
\end{eqnarray*}
 with transition function given by \( \Omega =\left( \begin{array}{cc}
 & 1\\
-1 & 
\end{array}\right)  \) (i.e. \( g_{(i+1)L}=\Omega g_{(i)L} \) with 
\( i\: \: mod\:4 \)) for \( g_{L} \)
and similarly for \( g_{R} \). If we do not want to use charts we have to use
singular functions as it happens with the Dirac monopole.

\subsection{Canonical formalism.}

Since we want to to discuss canonical transformations from interacting fields
to Wakimoto ones we need to set up the canonical formalism. This is easily done
and we find the momenta 

\begin{eqnarray*}
{\cal P} & = & \frac{|k|}{2\pi }\dot{\rho }\: \: \: \: {\cal P}_{+}=\frac{|k|}{2\pi }e^{2\rho }\partial _{+}x^{-}\: \: \: \: {\cal P}_{-}=\frac{|k|}{2\pi }e^{2\rho }\partial _{-}x^{+}
\end{eqnarray*}
along with the classical hamiltonian

\[
{\cal H}=\frac{\pi }{|k|}{\cal P}^{2}+\frac{|k|}{4\pi }\rho '^{2}+\frac{4\pi }{|k|}e^{-2\rho }{\cal P}_{+}{\cal P}_{-}-{\cal P}_{-}x'^{-}+{\cal P}_{+}x'^{+}\]
Moreover we can write the non vanishing Poisson brackets as
\begin{eqnarray}
\label{can-Poi-bra}
\left\{ x^{+}(\sigma ),{\cal P}_{+}(\sigma ')\right\} 
&=&\left\{ x^{-}(\sigma ),{\cal P}_{-}(\sigma ')\right\} 
=\delta (\sigma -\sigma ')
\nonumber\\
\left\{ \rho (\sigma ),{\cal P}(\sigma ')\right\} 
&=&\delta (\sigma -\sigma ')
\end{eqnarray}
Obviously this expressions are not very useful because we cannot use them to
deduce the commutation relations between the ``oscillators'' \( a,b,c \)
and \( \overline{a},\overline{b},\overline{c} \) due to the highly non linear
way they enter the expressions for \( x^{\pm }, \) \( \rho  \), explicitly
\begin{eqnarray}
{\cal P}_{+} & = & \frac{|k|}{2\pi }e^{2c}\partial _{+}b\: \: \: \: \: \: {\cal P}_{-}=\frac{|k|}{2\pi }e^{2\overline{c}}\partial _{-}\overline{b}\label{P+} 
\end{eqnarray}

\subsection{The KM algebra and the energy-momentum tensor.}

From the standard classical expression for the left/right KM currents \( J=|k|g^{-1}\partial _{+}g \)
(\( \overline{J}=|k|\partial _{-}gg^{-1} \)) we can compute the classical KM
currents which read
\begin{eqnarray}
J^{-} & = & |k|\: e^{2\rho }\partial _{+}x^{-}
\\
J^{3}&=& |k|\: \left( \partial _{+}\rho -x^{+}e^{2\rho }\partial_{+}x^{-}\right) 
\label{J-} \\
J^{+} & = & |k|\: \left( \partial _{+}x^{+}
+2x^{+}\partial _{+}\rho -x^{+2}e^{2\rho }\partial _{+}x^{-}\right) 
\nonumber\\
{}\label{J+} 
\end{eqnarray}
and 
\begin{eqnarray}
\overline{J}^{-} & = & |k|\: \left( \partial _{-}x^{-}
+2x^{-}\partial _{-}\rho -x^{-2}e^{2\rho }\partial _{-}x^{+}\right) 
\nonumber\\
{}\label{Jbar-} \\
\overline{J}^{3} & = & |k|\: \left( \partial _{-}\rho -x^{-}e^{2\rho
}\partial _{-}x^{+}\right) 
\\
\overline{J}^{+} & = & |k|\: e^{2\rho }\partial _{-}x^{+}\label{Jbar3} 
\end{eqnarray}
The previous currents can be rewritten in the canonical formalism in the following
way
\begin{eqnarray}
J^{-} & = & 2\pi {\cal P}_{+}
\nonumber\\
J^{3}&=&\frac{|k|}{2}\left( \frac{2\pi }{|k|}{\cal P}+\rho '\right) -2\pi x^{+}{\cal P}_{+}\label{j-} \\
J^{+} & = & |k|x'^{+}+|k|x^{+}\left( \frac{2\pi }{|k|}{\cal P}
+\rho'\right) 
\nonumber\\
&&-2\pi x^{+2}{\cal P}_{+}+2\pi e^{-2\rho }{\cal P}_{-}
\nonumber\\
{}\label{j+} 
\end{eqnarray}

\begin{eqnarray}
\overline{J}^{-} & = & -|k|x'^{-}+|k|x^{-}\left( \frac{2\pi} 
{|k|}{\cal P}-\rho '\right) 
\nonumber\\
&&-2\pi x^{-2}{\cal P}_{-}+2\pi e^{-2\rho }{\cal P}_{+}\label{jbar-} \\
\overline{J}^{3} & = & \frac{|k|}{2}\left( \frac{2\pi }{|k|}{\cal
P}-\rho '\right) -2\pi x^{-}{\cal P}_{-}
\nonumber\\
\overline{J}^{+} &=& 2\pi {\cal P}_{-}\label{jbar3} 
\end{eqnarray}
while the momentum-energy tensor reads
\begin{eqnarray*}
T_{++} & = & \frac{|k|}{4}\left( \frac{2\pi }{|k|}{\cal P}+\rho
'\right) ^{2}
+\frac{4\pi ^{2}}{|k|}e^{-2\rho }{\cal P}_{+}{\cal P}_{-}
\nonumber\\
&&+2\pi {\cal P}_{+}x'^{+}\\
T_{--} & = & \frac{|k|}{4}\left( \frac{2\pi }{|k|}{\cal P}-\rho
'\right) ^{2}
+\frac{4\pi ^{2}}{|k|}e^{-2\rho }{\cal P}_{+}{\cal P}_{-}
\nonumber\\
&&-2\pi {\cal P}_{-}x'^{-}
\end{eqnarray*}
It is then an easy matter to verify that they satisfy the following classical
Virasoro (with vanishing central charge) 
\begin{eqnarray}
\left\{ T_{++}(\sigma )\: ,\: T_{++}(\sigma ')\right\}  &  = &  
\nonumber\\
2\pi \: \left[ T_{++}(\sigma )+T_{++}(\sigma ')\right] \: \partial
_{\sigma }\delta (\sigma -\sigma ')
\nonumber\\
-2\pi \: \frac{c}{12}\: \partial ^{3}_{\sigma }
\delta (\sigma -\sigma ')\nonumber \label{Vir-class} \\
 &  & \\
\left\{ T_{--}(\sigma )\: ,\: T_{--}(\sigma ')\right\}  & = & 
\nonumber\\
-2\pi \: \left[ T_{--}(\sigma )+T_{--}(\sigma ')\right] \: 
\partial _{\sigma }\delta (\sigma -\sigma ')
\nonumber\\
+2\pi \: \frac{c}{12}\: \partial ^{3}_{\sigma }
\delta (\sigma -\sigma ')\nonumber \label{Vir-bar-clas} 
\end{eqnarray}
with \( c=0 \) and Kac-Moody algebra (of level \( |k| \) )
\begin{eqnarray}
\left\{ J^{a}(\sigma ),J^{b}(\sigma ')\right\}  & = & 
2\pi \: \epsilon ^{ab}_{.\: .\: c}\: J^{c}\: \delta (\sigma -\sigma ')
\nonumber\\
&&+\pi |k|\: \eta ^{ab}\: 
\delta ' (\sigma -\sigma ')\label{KM-clas} 
\\
\left\{ \overline{J}^{a}(\sigma ),\overline{J}^{b}(\sigma ')\right\}
& = & 
-2\pi \: \epsilon ^{ab}_{.\: .\: c}\: \overline{J}^{c}\: 
\delta (\sigma -\sigma ')
\nonumber\\
&&-\pi |k|\: \eta ^{ab}\: 
\delta ' (\sigma -\sigma ')\label{KM-bar-clas} 
\end{eqnarray}
 It is not difficult to check the classical Sugawara construction, i.e. \( T=\frac{1}{|k|}\eta _{ab}J^{a}J^{b} \).

\subsection{The classical canonical transformation to the Wakimoto fields.}

We are now ready to discuss the classical Wakimoto canonical fields. In order
to get a hint on how they are related to our starting canonical variables we
evaluate \( T \) on the general solution of the equations of motion and we
get
\[
T_{++}=|k|(\partial _{+}c)^{2}+|k|e^{2c}\partial _{+}b\: \partial _{+}a\]
Remembering the value of \( {\cal P}_{+} \) given in eq. (\ref{P+}) the previous
expression suggests that \( c \) has something to do with a ``free'' field
while \( a \) could be proportional to the field canonically conjugate to \( {\cal P}_{+} \).
Starting from this observation it is not difficult to show that 
\begin{eqnarray}
F=\sqrt{2|k|}(c+\overline{c}) & \label{F} \\
\beta =2\pi i{\cal P}_{+},\: \gamma =a
\\
\overline{\beta }=2\pi i{\cal P}_{-},\: \overline{\gamma }=\overline{a} & \label{bet-gam} 
\end{eqnarray}
which satisfy the following canonical Poisson brackets 
\begin{eqnarray}
\left\{ F(\sigma )\: ,\: \dot{F}(\sigma ')\right\} &=& 4\pi \: 
\delta (\sigma -\sigma ')
\\
\left\{ \beta (\sigma )\: ,\: \gamma (\sigma ')\right\}  
&=&-2\pi i\: \delta (\sigma -\sigma ')
\\
\left\{ \overline{\beta }(\sigma )\: ,\: \overline{\gamma }(\sigma
')\right\} 
&=&-2\pi i\: \delta (\sigma -\sigma ')\label{F-com-rel} 
\end{eqnarray}
reproduce eq.s (\ref{can-Poi-bra}).

We can now use these new canonical variables to express the classical energy
momentum tensor
\[
T_{++}=\frac{1}{2}\left( \partial _{+}F\right) ^{2}-i\beta \partial _{+}\gamma \]
and the classical left \( sl(2,R) \) KM generators 
\begin{eqnarray}
J^{-} & = & -i\beta \: \: \: \: \: \: 
J^{3}=\sqrt{\frac{|k|}{2}}\partial _{+}F+i\beta \gamma \label{clas-wak--} \\
J^{+} & = & |k|\partial _{+}\gamma +\sqrt{2|k|}\gamma \partial _{+}F+i\beta \gamma ^{2}\label{clas-wak-+} 
\end{eqnarray}
and the classical right \( sl(2,R) \) ones
\begin{eqnarray}
\overline{J}^{+} & = & -i\overline{\beta }\: \: \: \:
\overline{J}^{3}=-\sqrt{\frac{|k|}{2}}\partial _{-}F+i\overline{\beta }\overline{\gamma }\label{clas-wak--bar} \\
\overline{J}^{-} & = & |k|\partial _{-}\overline{\gamma }-\sqrt{2|k|}\overline{\gamma }\partial _{-}F+i\overline{\beta }\overline{\gamma }^{2}\label{clas-wak-+bar} 
\end{eqnarray}

\section{Monodromy matrices and boundary conditions.}

Which are the boundary conditions to be imposed, here we follow the approach
of (\cite{GerNev})? Usually and naively we would take \( g_{R},g_{L}\in SL(2,R) \)
but in this case proceeding in this way we would get some very ugly results
or miss two different physical sectors as we are going to explain, we take therefore
\( g_{R},g_{L}\in SL(2,C) \) with the further condition dictated by reality
of \( g \)
\begin{equation}
\label{g-cc}
g_{L}^{*}=g_{CC}g_{L}\: \: \: g_{R}^{*}=g_{R}g_{CC}^{-1}\: \: \: \: g_{CC}\in SL(2,C).
\end{equation}

We can now ask which is the most general boundary conditions we can impose on
\( g_{L,R} \) compatible with eq.s (\ref{bc-rho}-\ref{bc-x+}). The simplest
answer is the periodicity in \( g \) which can be achieved by imposing the
following boundary conditions
\begin{eqnarray}
g_{L}(\xi ^{+}+2\pi ) & = & g_{P}g_{L}(\xi ^{+})
\nonumber\\
g_{R}(\xi ^{-}-2\pi ) &=&g_{R}(\xi ^{-})g_{P}^{-1}\label{g-per-0} 
\end{eqnarray}

A less obvious answer which is nevertheless allowed by the boundary conditions
is 
\begin{eqnarray}
g_{L}(\xi ^{+}+2\pi ) & = & g_{P}g_{L}(\xi ^{+})g_{L0}
\nonumber\\
g_{R}(\xi ^{-}-2\pi ) &=& g_{R0}g_{R}(\xi ^{-})g_{P}^{-T}
\label{g-per} 
\end{eqnarray}
where \( g_{L0}=\left( \begin{array}{cc}
1 & 2\pi w\\
0 & 1
\end{array}\right)  \) and \( g_{R0}=\left( \begin{array}{cc}
1 & 2\pi \overline{w}\\
0 & 1
\end{array}\right)  \) as it can be checked from (\ref{bc-rho}-\ref{bc-x+}). From the explicit form
of the Wakimoto fields it turns out that both \( w \) and \( \overline{w} \)
have vanishing Poisson brackets with all the fields: this is not strange as
it can appear because the same happens in the flat limit.

Obviously eq. (\ref{g-cc}) has to be compatible with eq. (\ref{g-per}), i.e.
\( g_{CC} \) and \( g_{P} \) have to satisfy
\begin{equation}
\label{g0-gp-cons}
g_{CC}g_{P}=g_{P}^{*}g_{CC}
\end{equation}
As far as the periodicity is concerned there are three different equivalence
classes: \( g_{P} \) can be either hyperbolic, parabolic or elliptic. Such
classes correspond to tachionic, massless and massive string excitations in
the flat limit \( k\rightarrow \infty  \) .

\subsubsection{Hyperbolic sector.}

Let us start with the hyperbolic class is the most natural with the coordinates
associated with the Gauss decomposition (\ref{g-gauss}). This case can be described,
up to conjugation by a constant element, by taking 
\[
g_{P}=\left( \begin{array}{cc}
e^{2\pi p} & \\
 & e^{-2\pi p}
\end{array}\right) \: \: \: \: \: p>0\]
where the constraint \( p>0 \) is due to the symmetry \( g_{L}\rightarrow \Omega g_{L} \)
with \( \Omega =\left( \begin{array}{cc}
 & 1\\
-1 & 
\end{array}\right)  \). This form of the periodicity matrix does not fix completely the invariance
(\ref{gauge-inv}) which can, for example, be generically fixed by the further
constraint \( \overline{b}(0)=b(0) \); analogous considerations apply for the
other sectors. This \( g_{P} \) is compatible with \( g_{CC}=1 \). 

We can write the explicit expansions of the functions entering the general solution
with the given boundary conditions as

\begin{eqnarray}
a=w\xi ^{+}+a_{periodic},  b=e^{2p\xi ^{+}}b_{periodic},  c=p\xi
^{+}+c_{periodic}
\Rightarrow 
\nonumber\\
F=p\tau +F_{periodic},\beta =\beta _{periodic},\gamma =w\xi ^{+}+\gamma _{periodic}\nonumber \label{field-exp-hyper} 
\end{eqnarray}
 similarly for the barred quantities with \( \overline{p}=p \) but with \( \overline{w} \)
independent of \( w \). The presence of these two constants \( w \) and \( \overline{w} \)
is allowed by the equality of the variations of \( x^{\pm } \) at \( \sigma =0,2\pi  \)
in particular setting \( w=-\overline{w}=Rn \) (\( n\in Z \)) is equivalent
to compactify the \( x^{1} \) with radius \( R \), i.e. to choose an extremal
BH as background (\cite{MalStr}). It is important to notice that both \( w \)
and \( \overline{w} \) have vanishing Poisson brackets with everything as it
can be verified from the free field representation. Analogous considerations
apply to the other sectors.

It is interesting to notice that this is the only sector considered in (\cite{Bars},\cite{Satoh})
as it can be seen from the \( F \) expansion.

\subsubsection{Elliptic sector.}

Let us now consider the elliptic case which describes massive excitations in
the flat limit where
\[
g_{P}=\left( \begin{array}{cc}
\cos 2\pi p & \sin 2\pi p\\
\sin 2\pi p & \cos 2\pi p
\end{array}\right) \approx \left( \begin{array}{cc}
e^{i2\pi p} & \\
 & e^{-i2\pi p}
\end{array}\right) \: \: \: \: \: 0<p<\frac{1}{2}\]
where the first expression is the natural one when restricting the attention
to real \( g_{L,R} \) , i.e. when \( g_{CC}=1 \) while the second one is the
most natural when considering complex \( g_{L,R} \) with \( g_{CC}=\pm \left( \begin{array}{cc}
 & i\\
i & 
\end{array}\right)  \).

If we insist to use the real fields we get strange and ugly boundary conditions
such as
\[
c(\xi ^{+}+2\pi )=c(\xi ^{+})+\log \left( \cos 2\pi p+b(\xi ^{+})\sin 2\pi p\right) \]
which propagate to strange, non free boundary conditions for the Wakimoto fields
while using complex fields we get nice and free boundary conditions since the
complex fields have analogous expansion as (\ref{field-exp-hyper}) with the
substitution \( p\rightarrow i(p+k) \) (where \( k\in Z \)) but it obliges
us to impose the constraints
\[
a^{*}=a+\frac{1}{be^{2c}},\: \: \: b^{*}=\frac{1}{b},\: \: \:
c^{*}=c+\log \left( \pm ib\right) \]
They can be imposed in a better way by requiring the reality of the KM
currents.

Notice that if we restrict the momentum \( p \) to the first Block wave the
constraint \( 0<p<\frac{1}{2} \) implies \( 0<j=p_{F}<\frac{k}{2} \) which
is equivalent to the unitary spectrum truncation at the quantum level.

\subsubsection{Parabolic sector.}

Let us now consider the parabolic case where
\[
g_{P}=\left( \begin{array}{cc}
1 & 0 \\
2\pi p & 1
\end{array}\right) \: \: \: \: \: p\in R\]
and \( g_{CC}=1 \), then the fields in the 0th patch can be expanded as
\begin{eqnarray*}
a=w\xi ^{+}+a_{periodic},\: \: b=p\xi ^{+}+b_{periodic},
\nonumber\\
c=c_{periodic}\Rightarrow 
F=F_{periodic},
\nonumber\\
\beta =\beta _{periodic},\gamma =\gamma _{periodic}
\end{eqnarray*}
However life is not so easy since the fields in the 1st patch satisfy
\begin{eqnarray*}
a_{(1)}(\xi ^{+}+2\pi ) & = & 
a_{(1)}(\xi ^{+})-\frac{e^{-2c_{(1)}}}{1-2\pi p b_{(1)}(\xi^{+})}
\nonumber\\
\frac{1}{b_{(1)}(\xi ^{+}+2\pi )}&=&\frac{1}{b_{(1)}(\xi ^{+})}-2\pi p\\
c_{(1)}(\xi ^{+}+2\pi ) & = & c_{(1)}(\xi ^{+})
+\log \left( 1-2\pi p b_{(1)}(\xi ^{+})\right) 
\end{eqnarray*}
There is apparently not an easy way to impose these constraints on the
Wakimoto fields however it is necessary to use patches in this sector. 
A comment is now necessary in order to explain why it is necessary to work with
two patches in order the canonical transformation work fine, exactly as it happens
for the Liouville case. . If \( p_{F}\neq 0 \) we can express \( b \) using
the new canonical variables \( F \) and \( \beta  \) but if we try to solve
for \( b \) when \( p_{F}=0 \) then we cannot recover \( b_{0} \) (the constant
mode of \( b_{periodic} \) ) from the expression for \( \beta  \). This can
clearly be avoided if we use two patches.

\section{Conclusions.}

We have shown that a free fields approach to string propagating on \( AdS_{3} \)
requires a lot of attention and that we must work with different charts, as
done in (\cite{GerNev}) for the Liouville theory, if we want to treat the parabolic
sector correctly. As byproducts of this analysis we have shown that the spectrum
proposed by (\cite{Bars},\cite{Satoh}) is unnaturally truncated to the hyperbolic
sector and that it is possible to describe a string propagating on an extremal
BH background without much effort.

\end{document}